\documentclass[fleqn,usenatbib]{mnras}

\usepackage[T1]{fontenc}
\usepackage[utf8]{inputenc}
\usepackage{ae,aecompl}

\usepackage{graphicx}	% Including figure files
\usepackage{amsmath}	% Advanced maths commands
\usepackage{amssymb}	% Extra maths symbols

\title[Extinct radio pulsars]{Extinct radio pulsars as a source of subrelativistic positrons}

\author[Ya. N. Istomin et al.]{
Ya. N. Istomin,$^{1}$\thanks{E-mail: istomin@lpi.ru}
D. O. Chernyshov,$^{1}$\thanks{E-mail: chernyshov@lpi.ru}
and D. N. Sob'yanin$^{1}$\thanks{E-mail: sobyanin@lpi.ru}
\\
$^{1}$P. N. Lebedev Physical Institute of the Russian Academy of Sciences, Leninskii Prospekt 53, Moscow 119991, Russia}

\date{Accepted 2020 July 16. Received 2020 July 16; in original form 2020 April 08}

\pubyear{2020}

\begin{document}
\label{firstpage}
\pagerange{\pageref{firstpage}--\pageref{lastpage}}
\maketitle

\begin{abstract}
Extinct radio pulsars, in which stationary, self-sustaining generation of a relativistic electron-positron plasma becomes impossible when rotation brakes down, can be sources of a subrelativistic flux of positrons and electrons. We assume that the observed excess of positrons in the bulge and the disc of the Galaxy is associated with these old neutron stars. The production of pairs in their magnetospheres occurs due to one-photon absorption of gamma quanta of the Galactic and extragalactic backgrounds. The cascade process of plasma production leads to the flux of positrons escaping from the open magnetosphere $\simeq 3 \times 10^{34} \text{ s}^{-1}$. The total flux of positrons from all old Galactic neutron stars with rotational periods $1.5 < P < 35$~s is $\simeq 3 \times 10^{43} \text{ s}^{-1} $. The energy of positrons is less than $\simeq 10$~MeV. The estimated characteristics satisfy the requirements for the positron source responsible for the 511-keV Galactic annihilation line.	
\end{abstract}


%%%%%%%%%%%%%%%%%%%%%%%%%%%%%%%%%%%%%%%%%%%%%%%%%%

%%%%%%%%%%%%%%%%% BODY OF PAPER %%%%%%%%%%%%%%%%%%

\section{Introduction}

One of the most famous puzzles related to the problem of the acceleration and propagation of Galactic leptons is the origin of positrons responsible for the annihilation emission from the Galactic centre. This emission consists of the line of electron-positron annihilation with a characteristic energy of 511 keV and also of continuum emission due to 3-photon annihilation. The emission from the Galactic centre was the first gamma-ray line discovered outside the Solar system. It was detected using the balloon-borne detectors \citep{john72} and afterward was excessively studied by other experiments \citep[see, e.g., the review of][]{pranz11}. The most detailed observations were made using the SPI spectrometer located on the International Gamma-Ray Observatory ({\it INTEGRAL}) \citep{chur2005,kn2005,weid2006}.

A recent analysis of {\it INTEGRAL}/SPI data performed by \citet{sieg16} demonstrated that the actual bulge-to-disc ratio for the annihilation emission may be significantly lower than it was stated earlier. The reason for this drastic change was taking into account an additional spatial component of the annihilation emission which coincides with the thick disc of the Galaxy. Therefore, it was concluded that sources of positrons should somehow be related to old stars, which are abundant both in the Galactic bulge and in the thick disc. The annihilation rates were determined as $2\times 10^{43}$~s$^{-1}$ in the bulge and $3\times 10^{43}$~s$^{-1}$ in the thick disc \citep{sieg16}.

It is worth noting that independent observations by the balloon-borne Compton Spectrometer and Imager (COSI) showed a different spatial structure of the annihilation emission \citep{cosi19}. The annihilation flux from the Galactic bulge is, however, consistent with the SPI measurements \citep{sieg_cosi}. As pointed by \citet{sieg_cosi}, the discrepancy in the spatial structure may result from an inability of a coded-mask instrument such as SPI to extract large-scale emission, e.g., coming from the Galactic halo.

A complete model of the annihilation emission from the Galaxy should explain both the spatial morphology and the spectral properties of the emission. Since the spatial profile of the emission suggested by \citet{sieg16} is more or less consistent with the stellar population, it is more natural to assume that positrons do not travel far away from their production sites and to search for the sources with the spatial distribution similar to that of the annihilation emission. Indeed, the simulations of \citet{jean09,martin12,alex2014} show that it is very unlikely for positrons to travel more than $\sim 200$~pc away from their sources \citep[see the review of][for more details]{Panther2018a}.

Spectral properties of the emission require positrons to annihilate in a warm moderately ionized medium \citep{chur2005,jean06,chur11} being cooled down to the energies in the order of thermal energy. Since positrons in most cases are born almost relativistic, this energy excess should somehow be dissipated during their lifetime.

Mildly relativistic positrons lose energy mainly due to collisions with free or bound electrons, i.e., due to Coulomb or ionization losses \citep{pranz11}. However, exactly the same collisions also lead to annihilation and bremsstrahlung producing gamma-ray emission above 511 keV with hard spectrum. Since the annihilation emission does not correlate with the Galactic background, it is possible to extract the annihilation component and use it to restrict the initial energy of positrons \citep{ahar91}. A comparison between theoretical predictions and {\it INTERGAL} and COMPTEL data \citep{beacom} as well as a combination of COMPTEL, IBIS, and EGRET data \citep{sizun} showed that the maximum energy of annihilating positrons cannot exceed $1-3$ MeV for neutral medium and $3 - 10$ MeV for fully ionized medium.

This restriction can be relaxed if there is a collisionless mechanism of energy dissipation that dominates over collisional losses for the energies above $3-10$ MeV, such as adiabatic \citep{Panther2018b} or synchrotron \citep{chern10} losses. The first model imposes certain requirements on the background medium, while the second requires the magnetic field in the order of several mG, i.e., unrealistically strong.

Taking into account the above arguments, we summarise the requirements for the sources of positrons in the following way:
\begin{itemize}
\item The sources should be able to produce more than $2\times 10^{43}$ positrons per second in the Galactic bulge and more than $3\times 10^{43}$ positrons per second in the thick disc
\item The maximum energy of positrons produced by the sources should not exceed $3 - 10$~MeV
\item It is likely that the spatial distribution of the sources follows that of the old stellar population
\end{itemize}

Two classes of sources are presently considered as the main candidates. The first one involves $\beta^+$ decay of unstable nuclei produced during supernova explosions. In the recent paper by \citet{Crocker17} it was shown that the special subclass of SNe Ia can produce enough ${}^{44}$Ti to almost completely explain the origin of the Galactic annihilation emission.

Another class of sources involves leptonic pairs produced by jets in low-mass X-ray binaries (LMXBs) and microquasars. This model is supported by the observed annihilation emission from V404 Cygni \citep{sieg16_quasar}. \citet{bartels18} showed that LMXBs can be responsible not only for the annihilating positrons in the Milky Way but also for the gamma-ray excess observed from the inner part of the Galaxy. However, the amount of positrons LXMBs can produce is still uncertain.

\begin{figure}
\includegraphics[width=\columnwidth]{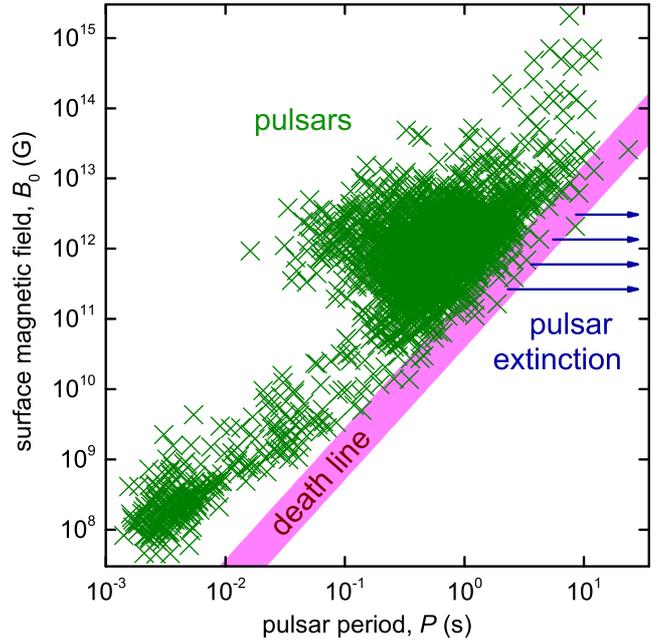}
\caption{The magnetic field at the stellar surface, $B_0$, against the period of rotation, $P$, for the observed pulsars ($P-B$ diagram, double logarithmic scale). Arrows schematically show trajectories of motion on the $P-B$ diagram for old neutron stars, extinct radio pulsars.}
\label{fig1}
\end{figure}

In this paper, we show that old neutron stars, which were radio pulsars, can be a source of Galactic positrons. Usual radio pulsars are magnetized, rapidly rotating neutron stars with radius $R \simeq 10^6 $~cm, and magnetization means that the star has an intrinsic magnetic moment $ \mu \simeq 10^{30}\text{ G}\,\text{cm}^3 $ corresponding to the typical magnetic field at the surface $ B_0 \simeq 10^{12}$~G. The rotational period $P$ of neutron stars varies from milliseconds to about ten seconds, and every pulsar can conveniently be presented by a point on the $P-B$ diagram (Fig.~\ref{fig1}). High magnetic fields of neutron stars and their rotation make them efficient generators of dense relativistic plasmas consisting of electrons and positrons and determining the observed activity \citep{IstominSobyanin2007,IstominSobyanin2011a}. The fact that neutron stars can emit leptons was confirmed by the observation of gamma-ray halos around the pulsars Geminga and B0656+14 \citep{milagro09, hawc17}.

The existence of both the magnetic field and the rotation of a neutron star is important for the plasma generation. In strong magnetic fields ($B\gtrsim10^8$~G, see Fig.~\ref{fig1}), one-photon pair production occurs (i.e. $\gamma\rightarrow e^+e^-$), and the created electron and positron are subsequently accelerated by the longitudinal electric field directed along the magnetic field and induced by rotation. The latter acceleration results in the radiation of new photons, and their absorption gives new pairs, thereby forming a cascade process of plasma multiplication. The stationary plasma generation near the stellar surface in the polar magnetosphere is maintained under the condition of a sufficiently strong magnetic field and a rather fast rotation, which defines the so-called death line on the $P-B$ diagram, $P \propto B_0^{ 8/15}$.

The possible contribution of regular and millisecond pulsars to the positron flux in the Galaxy was discussed previously \citep{wang06}, but the energy of positrons was found to be too high to explain the Galactic annihilation emission \citep{jager08, pranz11}. Meanwhile, electrons and positrons will have lower energies than those going from the usual pulsars if these are produced in the magnetospheres of less energetic neutron stars. Such neutron stars may have larger rotational periods but must have the magnetic fields of a typical pulsar because such fields are necessary for one-photon pair production to be effective. When rotation of the neutron star brakes and the energy of rotation is transmitted mainly to the plasma generation, the star approaches the boundary of `death' on the $P-B$ diagram and ceases to be a radio pulsar, now becoming an `extinct pulsar' (see the Fig.~\ref{fig1}). We show that extinct pulsars are able to produce enough $e^+e^-$ pairs to account for the annihilation emission. Besides, these sources, unlike active pulsars, produce particles with energy low enough to satisfy the requirements on the positron source responsible for the 511-keV Galactic annihilation line.

The paper is structured as follows. In section 2 we will discuss how an electron-positron plasma is generated in the magnetospheres of slowly rotating neutron stars, extinct radio pulsars. Section 3 determines the parameters of the plasma escaping from the open magnetosphere. Section 4 is devoted to conclusions that follow from the presented model of the origin of subrelativistic positrons in the Galaxy.
 		
\section{Slowly rotating magnetized neutron stars}

After the neutron star has slowed down, intersected the death line on the $P-B$ diagram, and become an extinct pulsar, production of $e^+e^-$ pairs cannot continue near the stellar surface. Instead, pair production occurs in the magnetosphere. The problem of appearance of the neutron star magnetosphere from a vacuum with a strong magnetic field is very complex \citep{IstominSobyanin2009,IstominSobyanin2010a,IstominSobyanin2010b}. The diffuse Galactic and isotropic extragalactic gamma radiation illuminates it and causes one-photon pair production \citep{IstominSobyanin2011a}. Estimates show that about $ 2 \times 10^{10} $ gamma-ray photons with energy higher than the threshold energy $ 2m_ec^2 $ can be absorbed in the magnetosphere per second \citep{IstominSobyanin2011b}. The produced electrons and positrons are accelerated by the longitudinal electric field present in the rotating magnetosphere and emit gamma quanta that in turn produce pairs.

Thus, the plasma multiplication cascade develops, and the dense electron-positron plasma, filling a narrow tube of a size of about 100~m in the transverse direction, expands with the speed of light both in the direction towards the stellar surface and outwards, so that after the absorption of one external gamma-ray photon a `lightning' forms, a lengthening and simultaneously
expanding plasma tube developing on a ms time-scale. The lightning going towards the surface enters the region where the magnetic and longitudinal electric fields increase and produces a huge number of electron-positron pairs up to $ 10^{28} $ per external photon \citep{IstominSobyanin2011b}. On the basis of this effect, \citet{IstominSobyanin2011c} constructed a model for the formation of rotating radio transients (RRATs) from slow rotating neutron stars with $ P \simeq 1-10$~s. The other part of the lightning spreads outwards, and if it is in the open polar region of the magnetosphere, consisting of the magnetic field lines going to infinity, the plasma flows into the surrounding space.

\begin{figure}
\includegraphics[width=\columnwidth]{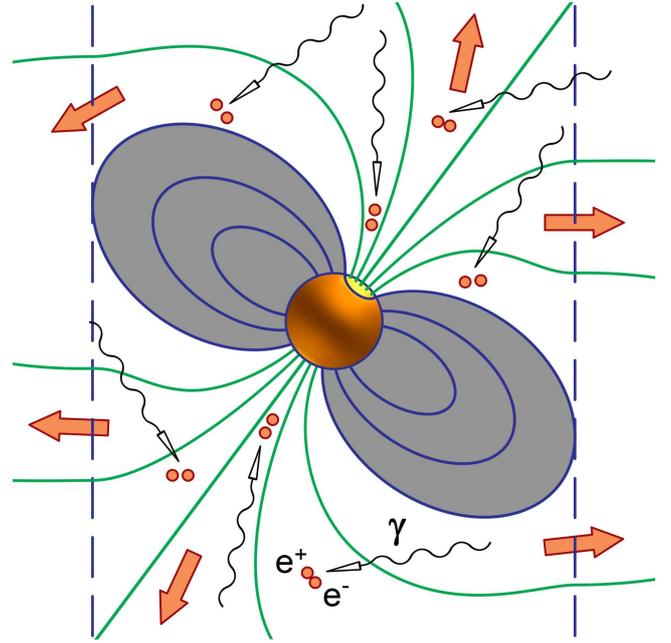}
\caption{Production of electrons and positrons in the open magnetosphere of a neutron star by high-energy photons from the cosmic gamma-ray background. The figure shows the neutron star (golden sphere), the closed magnetosphere (grey-shaded area), where magnetic field lines (blue) begin and end at the stellar surface, and the open magnetosphere, where magnetic field lines (green) emanate from the polar cap (yellow) and go through the light cylinder (blue dashed lines) to infinity. The gamma-ray photons (wavy arrows) are absorbed in the magnetosphere and converted to $e^+e^-$ pairs (orange circles). The produced plasma flows along open magnetic field lines out of the magnetosphere (wide orange arrows).}
\label{fig2}
\end{figure}

In the outer magnetosphere, the strength of the magnetic and longitudinal electric fields rapidly falls, the plasma multiplication cascade becomes less effective than in the inner magnetosphere, and the energies of electrons and positrons become not as large as in the case of the part of the lightning propagating towards the surface. Here, the production of pairs involved in the cascade of plasma multiplication is strongly non-local in nature. The mean free path of photons with respect to pair production, the length of particle acceleration, and the radius of curvature of magnetic field lines become comparable to the size of the magnetosphere. The scheme of the electron-positron pair production is shown in Fig.~\ref{fig2}. We will try to determine from general considerations the characteristics of the electron-positron plasma escaping from the magnetosphere, its flow, and characteristic energy.

\section{Plasma outflows: positron flux and energy}

The characteristic size $r_\text{L}=c/\Omega\sim10^{10}$~cm of the magnetosphere of a rotating neutron star, where $c$ is the speed of light and $\Omega =2\upi/P\sim1\text{ s}^{-1}$ is the rotation frequency (period $P\sim10$~s), exceeds greatly the stellar radius $R\simeq 10^6$~cm and corresponds to the distance at which plasma corotation breaks down and its magnetic confinement becomes impossible. It determines a cylindrical light surface $r=r_\text{L}$ which contains the closed magnetosphere and at which the corotation velocity $\Omega r$ formally equals $c$ (see Fig.~\ref{fig2}). The typical vertical size of the magnetosphere is in the order of its radius, and the energy flux for the outflow of an electron-positron plasma escaping from the magnetosphere is
\begin{equation}
\label{kineticEnergyFlux}
W=4\upi r_\text{L}^2 n\gamma m_e c^3,
\end{equation}
where $n$ is the plasma number density and $\gamma$ is the Lorentz factor. In what follows, it is convenient to characterize the plasma density by a dimensionless multiplicity $\lambda=n/n_\text{GJ}$, where $n_\text{GJ}$ is the Goldreich-Julian density \citep{GoldreichJulian1969}, which approximately corresponds to the minimum particle density necessary for an equilibrium magnetospheric charge density when the magnetic field is not significantly twisted \citep{Sobyanin2016}. At the magnetospheric boundary we have
\begin{equation}
n_\text{GJ}=\frac{\Omega B_0}{2\upi ce}\left(\frac{\Omega R}{c}\right)^3.
\end{equation}

In order to determine the energy flux $W$ \eqref{kineticEnergyFlux} through the parameters of the neutron star, we first turn to the dimensions of the quantities under consideration. The energy flux eventually goes from the rotational energy of the star and owes its existence to the rotating magnetic field frozen-in into the star, i.e., to the stellar magnetic moment $\mathbf{\mu}$. Since $\mathbf{\mu}$ is a rotating vector, $W$ cannot be proportional to the first degree of $\mu$; we have $W\propto\mu^2$ because the corresponding dimensions are $[W]=ML^2T^{-3}$ and $[\mu]=M^{1/2}L^{5/2}T^{-1}$, where $M$, $L$, and $T$ are dimensions of mass, length, and time, so that $[W]=[\mu^2]L^{-3}T^{-1}$. Since $R\ll r_\text{L}$, the characteristic length is $r_\text{L}=c/\Omega$, while the characteristic time is $\Omega^{-1}$. Thus,
\begin{equation}
\label{magneticEnergyFlux}
W=a\frac{\Omega^4 \mu^2}{c^3}=5.8\times10^{31}aB_{12}^2P^{-4}R_6^6\text{ erg}\,\text{s}^{-1},
\end{equation}
where $a\simeq1$ is a dimensionless factor. Note that $a=(2/3)\sin^2\chi$ for vacuum magnetic dipole losses and $a=i\cos^2\chi$ for current losses in a plasma-filled case, where $\chi$ is the inclination angle and $i$ is a dimensionless current flowing in the magnetosphere \citep{BeskinGurevichIstomin1984}, so that $a\simeq1$ exactly corresponds to an axisymmetric pulsar with a more general dependence in an inclined case \citep{Gruzinov2005,Gruzinov2006,Spitkovsky2006}.
It should be emphasized here that the three quantities $\Omega$, $c$, and $\mu$ entering \eqref{magneticEnergyFlux} and containing the necessary three dimensions $T$, $L$, and $M$ are not just arbitrary quantities but are the specific physical quantities that characterize the real parameters of a rotating magnetized neutron star -- the rotation frequency ($\Omega$), the magnetospheric size ($r_\text{L}=c/\Omega$), and the magnetic moment ($\mu$) -- but not their fractions or their arbitrary combinations. From these considerations it follows that the dimensionless factor $a$ is generally of order unity, $a\simeq 1$.

Equating \eqref{kineticEnergyFlux} and \eqref{magneticEnergyFlux} gives
\begin{equation}
\label{lambdaGamma}
\lambda\gamma=\frac{a\omega_c}{2\Omega}\left(\frac{\Omega R}{c}\right)^3=1.3\times10^7 a B_{12}P^{-2}R_6^3,
\end{equation}
where $\omega_\text{c}=eB_0/m_e c$ is the non-relativistic cyclotron electron frequency, $B_0$ is the magnetic field at the stellar surface, $B_{12} = B_0/10^{12}$~G and
$R_6=R/10^6$~cm are the dimensionless surface field and stellar radius. Equation \eqref{lambdaGamma} gives reasonable estimations for radio pulsars, $\lambda\sim10^4-10^5$ and $\gamma\sim10^2-10^3$ \citep{IstominSobyanin2007,TimokhinHarding2019}.

Another way to estimate the multiplicity is as follows: When the magnetosphere is filled with a plasma, the braking of stellar rotation can be brought about only by electric currents flowing in the magnetosphere and being closed at the stellar surface. The currents create the torque of electromagnetic forces braking the rotation. Since the magnetic field in the closed magnetosphere rests on a highly-conducting surface of the neutron star, every magnetic tube has the same electric potential equal to the stellar potential and the electric current can flow in the open magnetosphere only. Let us estimate the electric current flowing out of the magnetosphere. When magnetic field lines in the open part of the magnetosphere go beyond the light surface, electrons and positrons move with slightly different velocities $v_{e,p}=c(1-\gamma_{e,p}^{-2}/2)$ because the longitudinal electric field accelerates one type of particles and decelerates the other, so the resulting current is
\begin{equation}
I=\lambda\frac{B_0\Omega^2 R^3}{2c}|\gamma_e^{-2}-\gamma_p^{-2}|.
\end{equation}
We have $|\gamma_e^{-2}-\gamma_p^{-2}|=b\gamma^{-2}$ with a factor $b<1$. The work of electric field on current $I$ per unit time is the power being lost by the star,
$W=U I$, where $U$ is the potential difference generated by the rotating magnetic field in the open magnetosphere; the relation is analogous to that takes place in relativistic jets, where the observed intensity corresponds to the power released in the jet via unipolar operation of the central engine, a supermassive black hole with a surrounding accretion disc \citep{Sobyanin2017}. Using Faraday's law $U=-(\partial\Phi/\partial t)/c$, where $\Phi=2\upi r_\text{L}^2B_0(\Omega R/c)^3$ is the magnetic flux through the light surface, we arrive at
\begin{equation}
U=2\upi B_0\frac{\Omega^2R^3}{c^2},
\end{equation}
and from $U I=W$ we get $\upi b\lambda=a\gamma^2$. Combining this relation with~\eqref{lambdaGamma}, we may separately find $\gamma$ and $\lambda$,
\begin{equation}
\label{gamma}
\begin{split}
\gamma&=\left(\frac{\upi b}{2}\right)^{1/3}\left(\frac{\omega_c}{\Omega}\right)^{1/3}
\left(\frac{\Omega R}{c}\right)\\
&=3.5\times10^2 b^{1/3}B_{12}^{1/3}P^{-2/3}R_6,
\end{split}
\end{equation}
\begin{equation}
\label{lambda}
\begin{split}
\lambda&=\left(\frac{a^3}{4\upi b}\right)^{1/3}\left(\frac{\omega_c}{\Omega}\right)^{2/3}
\left(\frac{\Omega R}{c}\right)^2\\
&=4\times10^4\left(\frac{a}{b^{1/3}}\right)B_{12}^{2/3}P^{-4/3}R_6.
\end{split}
\end{equation}

Interestingly, the surface magnetic field $B_0$ appears in relations \eqref{gamma} and \eqref{lambda} in a combination $B_0(\Omega R/c)^3$, which is the magnetic field at the light surface, $B_\text{L}=\mu/r_\text{L}^3$. The latter field does not depend separately on the surface field $B_0$ or the stellar radius $R$ but is determined by the stellar magnetic moment $\mu$ and angular velocity $\Omega$ only. If we define the cyclotron frequency for electrons and positrons at the light surface, $\omega_c^\text{L}=eB_\text{L}/m_ec$, relations \eqref{gamma} and \eqref{lambda} take on a simple form
\begin{equation}
\label{simpleGamma}
\gamma\simeq \left(\frac{\omega_\text{c}^\text{L}}{\Omega}\right)^{1/3},
\end{equation}
\begin{equation}
\label{simpleLambda}
\lambda\simeq \left(\frac{\omega_\text{c}^\text{L}}{\Omega}\right)^{2/3}.
\end{equation}
The ratio $\omega_\text{c}^\text{L}/\Omega$ is the particle magnetization at the light surface. The sole dependence of particle Lorentz factor on magnetization presented in \eqref{simpleGamma} is a universal property of the systems in which the acceleration of charged particles in a magnetic field occurs. Particularly, it is impossible to accelerate a particle whose cyclotron frequency $\omega_\text{c}/\gamma$ becomes less than the characteristic frequency of a system under study (Hillas' criterion) \citep{Hillas1984}, so the maximum Lorentz factor is $\gamma_\text{max} = \omega_\text{c}/\Omega$. In magnetospheres of rotating massive black holes $\gamma=(\omega_\text{c}/\Omega)^{\beta}$, where the index $\beta<1$ may change from $1/3$ to $2/3$. In the works by \citet{Michel1969,Michel1974} $\beta$ appears to be $1/3$. \citet{IstominGunya2020} have found that direct acceleration of protons by an electric field transmitting rotation from the black hole to the magnetospheric plasma is more efficient, $\beta = 1/2$ and $\beta = 2/3$ depending on the toroidal magnetic field in the magnetosphere.

Relation \eqref{simpleGamma} can also be obtained from the following order-of-magnitude estimations: The main particle acceleration occurs near the light surface, where the electric field $E$ becomes comparable to the magnetic field $B_\text{L}$. A particle during its gyration departs from the initial magnetic field line by a distance $l$, and the electric field does a work on it, so that $\gamma mc^2=eEl$. Since $E\simeq B_\text{L}$, we have $l\simeq l_0=\gamma mc^2/eB_\text{L}$ and then should use the criterion $l=r_\text{c}$. The cyclotron radius $r_\text{c}=v/(\omega_\text{c}/\gamma)$ is not identically equal to $l_0$: firstly, the particle velocity $v$ is slightly less than the velocity of light, $v=c(1-1/2\gamma^2)$, and secondly, the effective magnetic field on the particle trajectory is slightly less than $B_\text{L}$ because of its decrease with distance, $B\simeq B_\text{L}(1-r_\text{c}/r_\text{L})\simeq B_\text{L}[1-\gamma(\Omega/\omega_\text{c}^\text{L})]$. We drop off insignificant factors because the magnetic field at the light surface deviates from the dipole structure; besides, the electric field may also differ from $B_\text{L}$ by the terms of order $\gamma^{-2}$, and the same is true for the difference between $l$ and $l_0$. With these reservations, we get from $l=r_\text{c}$
\begin{equation}
l=l\left(1-\frac{1}{\gamma^2}\right)\left(1+\gamma\frac{\Omega}{\omega_\text{c}^\text{L}}\right).
\end{equation}
Equating the small terms gives $\gamma^3\simeq\omega_\text{c}^\text{L}/\Omega$, which corresponds to~\eqref{simpleGamma}.

Now let us find the positron flux $L=W/2\gamma m_e c^2$ created by the neutron star,
\begin{equation}
\label{positronFlux}
L=1.3\times10^{35}ab^{-1/3} B_{12}^{5/3}P^{-10/3}R_6^5 \text{ s}^{-1}.
\end{equation}
The factor $b<1$ appears in relations \eqref{gamma}, \eqref{lambda}, and \eqref{positronFlux} in degree $1/3$, so we may put $b^{1/3}\simeq 1$.

While increasing period, radio pulsars pass through the death line (see Fig.~\ref{fig1}) and finally go away to the region of large periods, where the stationary generation of an electron-positron plasma without external influence becomes impossible. In fact, the major portion of radio pulsars has periods $P<P_0\simeq 1.5$~s. Though one observes slowly rotating pulsars with periods reaching $\simeq 10$~s, these are few (note a unique ultra-slow pulsar PSR J0250+5854 with $P=23.5$~s \citep{TanEtal2018}). Then operating under the influence of the flux of gamma quanta from the external cosmic background, neutron stars continue braking and generating an electron-positron plasma that flows out of the magnetosphere along open magnetic field lines. The evolution of rotational period is described by the relation $J\Omega{\dot\Omega}=-W$, where $J\simeq 10^{45}\text{ g}\,\text{cm}^2$ is the stellar moment of inertia; therefore,
\begin{equation}
\label{rotationalPeriod}
P=P_0\left(1+\frac{t}{\tau}\right)^{1/2},
\end{equation}
where
\begin{equation}
\tau=\frac{c^3 P_0^2 J}{8\upi^2 a B_0^2 R^6}=1.1\times10^7\frac{P_0^2}{a}\frac{J_{45}}{B_{12}^2 R_6^6}\text{  yr},
\end{equation}
with $J_{45}=J/10^{45}\text{ g}\,\text{cm}^2$ being the dimensionless stellar moment of inertia.

The magnetic field of isolated neutron stars virtually does not decay \citep{BhattacharyaSrinivasan1991,HartmanEtal1997,JohnstonKarastergiou2017} and exists in superconducting vortices in the stellar core, so when estimating the maximum rotational period $P_\text{max}$ with the help of~\eqref{rotationalPeriod}, we put the time $t$ equal to the Galactic age $t_\text{G}\simeq 1.3\times10^{10}$~yr,
\begin{equation}
\label{Pmax}
P_\text{max}=P_0\left(1+ 1.2\times10^3\frac{a}{P_0^2}\frac{B_{12}^2 R_6^6}{J_{45}}\right)^{1/2}\simeq 35
\frac{a^{1/2}B_{12} R_6^3}{J_{45}^{1/2}}\text{ s}.
\end{equation}
We see that the maximum period of extinct pulsars is virtually independent of the period $P_0$ with which intersection of the death line has occurred. All the neutron stars that became radio pulsars and were born at the moment of the Galaxy formation, now have period $P=P_\text{max}\simeq 35$~s.

Let us define a pulsar birthrate $\nu(t)$ via $dN=\nu(t)dt$, so that the total number of neutron stars is $N=\int_0^{t_\text{G}} \nu(t')dt'\simeq 10^9$ \citep{ArnettSchrammTruran1989,PernaEtal2003}. If one assumes that every supernova explosion leads to the birth of a radio pulsar, then the rate $\nu_0$ of supernova explosions in the Galaxy corresponds to the birthrate of neutron stars. Current estimations give $\nu_0\simeq 3\times10^{-2} \text{ yr}^{-1}$ \citep{TammannLofflerSchroder1994,DiehlEtal2006,AdamsEtal2013}, and were this quantity constant in time, the number of neutron stars would be $4\times10^8$. It can be, however, that in the past, when the Galaxy was young, the rate of supernova explosions was higher, and the number of old neutron stars is greater than that of young \citep{Crocker17}. An alternative is the early star formation in the Galactic bulge up to $6\times10^8$ \citep{Ofek2009}. In any case, it is difficult to determine the present period distribution of old neutron stars without knowledge of the history of pulsar formation. The period lies in the range from $P_0\simeq 1.5$~s to $P_\text{max}\simeq 35$~s and depends on the magnetic field $B_0$ (note the pulsar death line and equation~\eqref{Pmax}). To find the average value of the Lorentz factor of outflowing particles, we put $P=P_\text{max}$ because $\gamma\propto P^{-2/3}$ corresponds to an integral dependence with positive exponent $1/3$,
\begin{equation}
\label{averageGamma}
{\bar \gamma}\simeq 3.5\times10^2 B_{12}^{1/3}P_\text{max}^{-2/3}R_6=33\left(\frac{J_{45}}{a B_{12} R_6^3}\right)^{1/3}.
\end{equation}

It is worth noting that estimate \eqref{averageGamma} refers to the energy of electrons and positrons flowing directly out of the neutron star magnetosphere. Expanding into the interstellar medium, a subrelativistic plasma cools down, and its Lorentz factor drops when the plasma enters the surrounding gas. Let us estimate this value $\gamma_\text{f}$. The plasma expands freely up to a distance $l$ where its pressure becomes equal to the pressure $P_\text{g}$ of interstellar gas, $P_\text{g}=2\gamma_\text{f}nm_ec^2$. For the relativistic gas, the plasma density $n$ at the boundary of the cavity of size $l$ is related to the density at the boundary of the magnetosphere by $n=n_0(r_\text{L}/l)^2$. On the other hand, the work done by the plasma for supplantation of the gas from the cavity is $4\upi P_\text{g} l^3/3$ and goes from the energy of the plasma flow, $4\upi({\bar \gamma}-\gamma_\text{f})m_e c^2n_0r_\text{L}^2l$. As a result, ${\bar \gamma}-\gamma_\text{f}=2\gamma_\text{f}/3$, $\gamma_\text{f}=3{\bar \gamma}/5$. Thus, the energy of positrons interacting with the interstellar gas is
\begin{equation}
\label{Ep}
E_p=\gamma_\text{f} m_e c^2\simeq 20 m_e c^2=10\text{ MeV}.
\end{equation}
With account of \eqref{averageGamma}, ${\bar \gamma}\propto B_{12}^{-1/3}$, the value \eqref{Ep} may be considered an upper limit; e.g., for $B_{12}\simeq 10$ we have $E_p \simeq 5$~MeV.

To estimate the total number of positrons produced by old neutron stars, we should put $P=P_0$ in \eqref{positronFlux} because the strong dependence $P^{-10/3}$, with integral dependence being $P^{-7/3}$, implies the main contribution at the lower limit of periods, $P\simeq P_0$,
\begin{equation}
\begin{split}
L_\text{tot}&\simeq 1.3\times10^{35}aNB_{12}^{5/3}P_0^{-10/3}R_6^5\text{ s}^{-1}\\
&=3.4\times10^{43}aN_9B_{12}^{5/3}R_6^5 \text{ s}^{-1}.
\end{split}
\end{equation}
Here, $N=10^9 N_9$ is the number of old neutron stars, $N_9\simeq 1$.

At the end of this section, we will discuss a question whether it is possible to observe such old neutron stars, which give the necessary flux of subrelativistic positrons. As in radio pulsars, the flow of charged particles, electrons and positrons, albeit at low energies, must emit radio waves. Relativistic particles, moving with acceleration across the magnetic field due to curvature of magnetic field lines, generate electromagnetic waves. This radiation, named curvature radiation, can reach high powers due to the collective generation effect \citep{IstominPhilippovBeskin2012}. For radio pulsars, the efficiency of generating radio waves in the magnetosphere, the ratio of the energy flux of radio emission to the energy flux of relativistic particles, is $ \alpha_\text{r}\simeq 10^{- 6} $ \citep{BeskinGurevichIstomin1993}. In our case of old neutron stars, $ P \simeq 15$~s and $B_{12} \simeq 1 $, the energy flux of the particles emanating from the magnetosphere is $ W \simeq 10^{27 }\text{ erg}\,\text{s}^{-1}$. Thus, the power of radio emission from an old neutron star is $ W_r \simeq 10^{21}\text{ erg}\,\text{s}^{-1}$.

However, the emitted radio waves have significantly lower frequencies, $ \omega = c \gamma^3 / \rho_\text{c} $, where $ \rho_\text{c} \simeq r_\text{L} = cP / 2 \upi $ is the radius of curvature of the magnetic field lines in the far open magnetosphere. For the characteristic value of the Lorentz factor of electrons and positrons in the magnetosphere, $\gamma\simeq 30$, we have the value of the frequency of radiated radio waves, $f=P^{-1}\gamma^3\simeq 2$~kHz. Such waves do not penetrate through the ionosphere towards the Earth, but can be detected in space. If we assume that the nearest old neutron star is at a distance of $\simeq 10$~pc, then the observed energy flux density should be $\simeq 4$~Jy.

It should be noted that such kind of radiation is observed in the near space. This is the so-called terrestrial kilometric radiation that occurs in the lower polar magnetosphere of the Earth \citep{WuLee1979}. The observed frequencies are $f\simeq 5\times(10^4 - 10^5)$~Hz, the power is equal to $10^{13} - 10^{14}\text{ erg}\,\text{s}^{-1}$. Radiation is observed on remote spacecrafts, even on the orbit of the Moon. Similar radiation was detected from other planets of the Solar system which have their own magnetic field. Then the Jupiter emits decametric radio emission.

\section{Conclusions}

Old neutron stars considered in this paper generate positrons with much lower energy not exceeding 10~MeV.  The total flux of positrons produced by such stars, $ \simeq 3\times 10^{43}\text{ s}^{- 1}$, corresponds to the necessary flux in the bulge and in the thick disc of the Galaxy.
As for the spatial distribution of extinct radio pulsars, it may seem that they, as well as active radio pulsars, should be widely distributed over the distance $z$ to the Galactic plane, $|z|\simeq 300$~pc. However, over a long period of time comparable to the age of the Galaxy, neutron stars with high vertical speeds begin to oscillate in the vertical direction in the gravitational potential of the Galaxy. Calculations show that half period of oscillations is in the order of $4\times10^7$~yr \citep{BSMKS2006}, which on the one hand is greater than the characteristic age of radio pulsars, $ \simeq 10^6 -10^7$~yr, and on the other hand is much less than the age of the Galaxy, $\simeq 10^{10}$~yr, so old neutron stars cross many times the Galactic plane, where they lose their velocity due to friction \citep {ZK2016, WLSFMB2018}. Thus, their distribution is close to the distribution of old stars, and they satisfy all the requirements for the positron source responsible for the 511-keV Galactic annihilation line.

\section{Acknowledgments}
We would like to thank Fiona H. Panther for useful comments and suggestions. When plotting Fig.~\ref{fig1}, we used the data from the ATNF Pulsar Catalogue, version 1.63, http://www.atnf.csiro.au/research/pulsar/psrcat \citep{ATNF}.
This work was supported by Russian Foundation for Fundamental Research, grant number 20-02-00469. D.O.C.
is supported by the grant RFBR 18-02-00075.

%%%%%%%%%%%%%%%%%%%%%%%%%%%%%%%%%%%%%%%%%%%%%%%%%%



\bibliographystyle{mnras}
\providecommand{\noopsort}[1]{}\providecommand{\singleletter}[1]{#1}%

%%%%%%%%%%%%%%%%%%%%%%%%%%%%%%%%%%%%%%%%%%%%%%%%%%

% Don't change these lines
\bsp	% typesetting comment
\label{lastpage}
\end{document}